\documentclass{elsarticle}

\usepackage{hyperref}

\journal{Journal of Molecular Spectroscopy}

\usepackage{graphicx}

\begin{document}

\begin{frontmatter}

\title{Incomplete rotational cooling in a 22-pole ion trap}

\author{E. S. Endres}

\author{G. Egger}

\author{S. Lee}

\author{O. Lakhmanskaya}

\author{M. Simpson}

\author{R. Wester}

\ead{roland.wester@uibk.ac.at}

\address{Institut f\"ur Ionenphysik und Angewandte Physik, Universit\"at Innsbruck, Technikerstr. 25, 6020 Innsbruck, Austria}


\begin{abstract}
Cryogenic 22-pole ion traps have found many applications in ion-molecule reaction kinetics and in high resolution molecular spectroscopy. For most of these applications it is important to know the translational and internal temperatures of the trapped ions. Here, we present detailed rotational state thermometry measurements over an extended temperature range for the two ion/buffer gas systems OH$^-$/He and OD$^-$/HD with ion-to-neutral mass ratios of 4.25 and 6 respectively. The measured rotational temperatures show a termination of the thermalisation with the buffer gas around 25\,K, independent of mass ratio and confinement potential of the trap. Different possible explanations for this incomplete thermalisation have been investigated, among them the thermalisation of the buffer gas and the heating due to room temperature blackbody radiation and room temperature gas entering the trap.
\end{abstract}

\end{frontmatter}

\section{Introduction}

In recent years, collisional cooling of trapped ions by buffer gas, usually consisting of a cryogenic noble gas, has been established as a standard technique to cool molecular ions in multipole traps \cite{gerlich1995:ps}. This has opened up a diverse field of research on rate coefficients of ion-molecule reactions of astrochemical relevance \cite{endres2014:jpc,mulin2015:pccp}, cross sections for photodetachment of interstellar ions \cite{best2011:apj,kumar2013:apj} and high resolution molecular spectroscopy \cite{brunken_apjl_2014,jusko2014:prl,asvany2015:sci,lee2016:pra}. As sensitive spectroscopic probes laser-induced reactions \cite{schlemmer1999:ijm}, desorption of a tagged atom \cite{kamrath2011:jacs,heine2013:jacs}, and photodetachment \cite{lee2016:pra} have been implemented. Quantum state-selected reactions have been studied at low temperatures \cite{gerlich2013:jpc} and inelastic collision rate coefficients have become accessible \cite{hauser2015:natp}. Furthermore, buffer gas rotational cooling has been combined with sympathetic cooling of the translational motion via laser-cooled atomic ions \cite{hansen_nat_2014}.

For experiments with trapped molecular ions, precise knowledge of the translational temperature of the ions as well as of the internal state distribution is desirable. Translational temperatures of trapped ions can be determined via high resolution spectroscopy resolving Doppler profiles. These techniques require narrow bound-bound transitions and thus are only applicable if a suitable transition and an efficient probing scheme is available \cite{schlemmer1999:ijm,jusko2014:prl}. Several simulation studies of the thermalisation process in multipole ion traps have been carried out \cite{asvany2009:ijm,wester2009:jpb,holtkemeier_PRL_2016,lakhmanskaya2014:ijm}. Up to now experiments regularly find larger ion temperatures than predicted by simulations (see cf.\ \cite{jusko2014:prl}).

Here we present negative ion photodetachment measurements undertaken in order to investigate the thermalisation of the ions' rotational temperature and possible heating effects. This work is based on the rotational temperature diagnostics by near-threshold photodetachment that our group has demonstrated previously \cite{otto2013:pccp}. Precise internal state thermometry measurements of OH$^-$ in He and  OD$^-$ in HD have been carried out over a temperature range of more than 100\,K. Furthermore, the rotational temperature was measured for different settings of the trapping potential. The effect of room temperature gas and of blackbody radiation entering through the openings of the trap are discussed. Finally, an indirect measurement is presented to probe the thermalisation of the buffer gas with the trap temperature.

\section{Theoretical method}

The excess electron of a negative ion in its internal ground state is bound with the energy of the electron affinity of the neutral molecule. If the anion is in an excited rotational state the electron binding energy is correspondingly reduced. A photon whose energy exceeds this electron binding energy can detach the electron. The total photodetachment cross section near threshold can be written as \cite{otto2013:pccp}:
\begin{equation}
\sigma_{PD}(h \nu) \propto \sum_{J} \frac{p(J, T)}{g_J} \left[ \sum_{b}  I_{Jb} \cdot \Theta(h \nu - \epsilon_{Jb}) (h \nu - \epsilon_{Jb})^p \right].
\label{eq1}
\end{equation}
The outer sum is over all considered rotational states $J$ of the anion and the inner sum over all possible transitions of the respective $J$ level to different neutral states. $p(J,T)$ is the population of the rotational level $J$ which weights the cross sections from each state. $I_{Jb}$ are the H\"onl-London factors giving the relative transition intensities \cite{schulz1983,goldfarb2005:jcp}. $\Theta$ is the Heaviside function taking only transitions above threshold into account, $h \nu$ is the photon energy and $\epsilon_{Ji}$ is the energy threshold of the transition. $g_J = 2J+1$ is the degeneracy of the rotational level which is by definition included in the H\"onl-London factors and the $p(J,T)$. The degeneracy must not be considered twice and is therefore removed by explicitly dividing the population by $g_J$. Finally $p$ determines the shape of the individual cross sections.

For the photodetachment of OH$^-$ it has been shown that $p=0.28$ works well \cite{engelking1982}. In the experiment, photodetachment induces a loss rate for the trapped ions, $k_{PD}$, which is proportional to the photodetachment cross section of the anion $\sigma_{PD}$ and the laser power $P_L$,
\begin{equation}
k_{PD} \propto P_L \cdot \sigma_{PD}.
\label{eq2}
\end{equation}
Just as the population $p(J,T)$ depends on the rotational temperature, the photodetachment cross section $\sigma_{PD}(h \nu)$ also changes with temperature. By measuring the loss rate at different transition frequencies, the rotational temperature of anions can be determined using equation \ref{eq1}.

\section{Experimental setup and methods}
\label{exp:setup}

A description of the employed 22-pole radiofrequency ion trap setup can be found in Ref.\ \cite{best2011:apj,otto2013:pccp} and only a brief account is given below. The hydoxyl molecular anions are produced in a pulsed plasma discharge source from a helium-water mixture. A time of flight mass spectrometer, based on Wiley-McLaren configuration, is used for mass selection prior to loading the ions into the cryogenic 22-pole rf trap. The multipole trap creates a radial confinement potential by applying an alternating electric field on the rods. The longitudinal trapping is achieved via a constant voltage on the entrance and exit endcaps. The multipole trap features a large field free region in the radial center with a steep rising potential towards the rods. This allows an efficient trapping with little influence from perturbing external fields \cite{gerlich1992:adv}.

The trap is built inside a copper housing, which has two openings to load and unload ions. The copper housing acts as a thermal shield and is mounted on a closed-cycle refrigerator, which achieves temperatures down to about 9\,K. The temperature of this thermal shield, i.\ e.\ the temperature of the trap ($T_T$), can be controlled between 9\,K and room temperature using resistive heaters mounted at the bottom of the trap. This temperature is measured with two commercial temperature diodes with a precision below 100\,mK and a typical temperature difference between the top and the bottom of the trap of one percent. 

The buffer gas that is injected into the trap is expected to thermalise with the thermal shield within a few collisions and subsequently cool the trapped anions. In an ideal case the anions should reach the temperature of the buffer gas ($T_{bg}$) which is also equal to the temperature of the thermal shield. Due to the thermal shield and the low temperature, the buffer gas density inside the trap is higher than in the surrounding vacuum chamber. The density inside the trap is measured with a gas species-independent capacitive pressure gauge, which held at room temperature and attached to the trap via a gas tube. This yields the pressure $p_{out}$, from which the density $\rho$ inside the trap is calculated using
\begin{equation}
\rho = \frac{p_{out} \cdot \alpha}{k_B \cdot \sqrt{T_{out} \cdot T_T}},
\label{eq4}
\end{equation}
based on the ideal gas approximation in the low-density molecular flow regime. $\alpha$ is the density enhancement factor that depends on the conductance through the mechanical openings of the ion trap, and $k_B$ is the Boltzmann factor. $T_{out}$ is the ambient temperature of the setup.

For the measurements, the 22-pole trap is loaded with a packet of OH$^-$ ions from the source. Buffer gas densities around 10$^{13}$/cm$^3$ are applied, which guarantees several thousand elastic collisions and several hundred rotationally inelastic collisions per second \cite{cote2000:pra,hauser2015:natp}. Therefore steady-state conditions are reached within several milliseconds. In the trap the ions are exposed to the photodetachment laser and detected on a microchannel plate detector. Power-normalised photodetachment loss rates near threshold are obtained by repeating the measurements for different exposure times, ranging between few hundred milliseconds and several seconds, and for different laser frequencies. These rates are fitted to the detachment cross section described by equation \ref{eq1}, which yields the rotational temperature of the trapped ions \cite{otto2013:pccp}.

\section{Results}

\begin{figure}
\includegraphics[width= \textwidth]{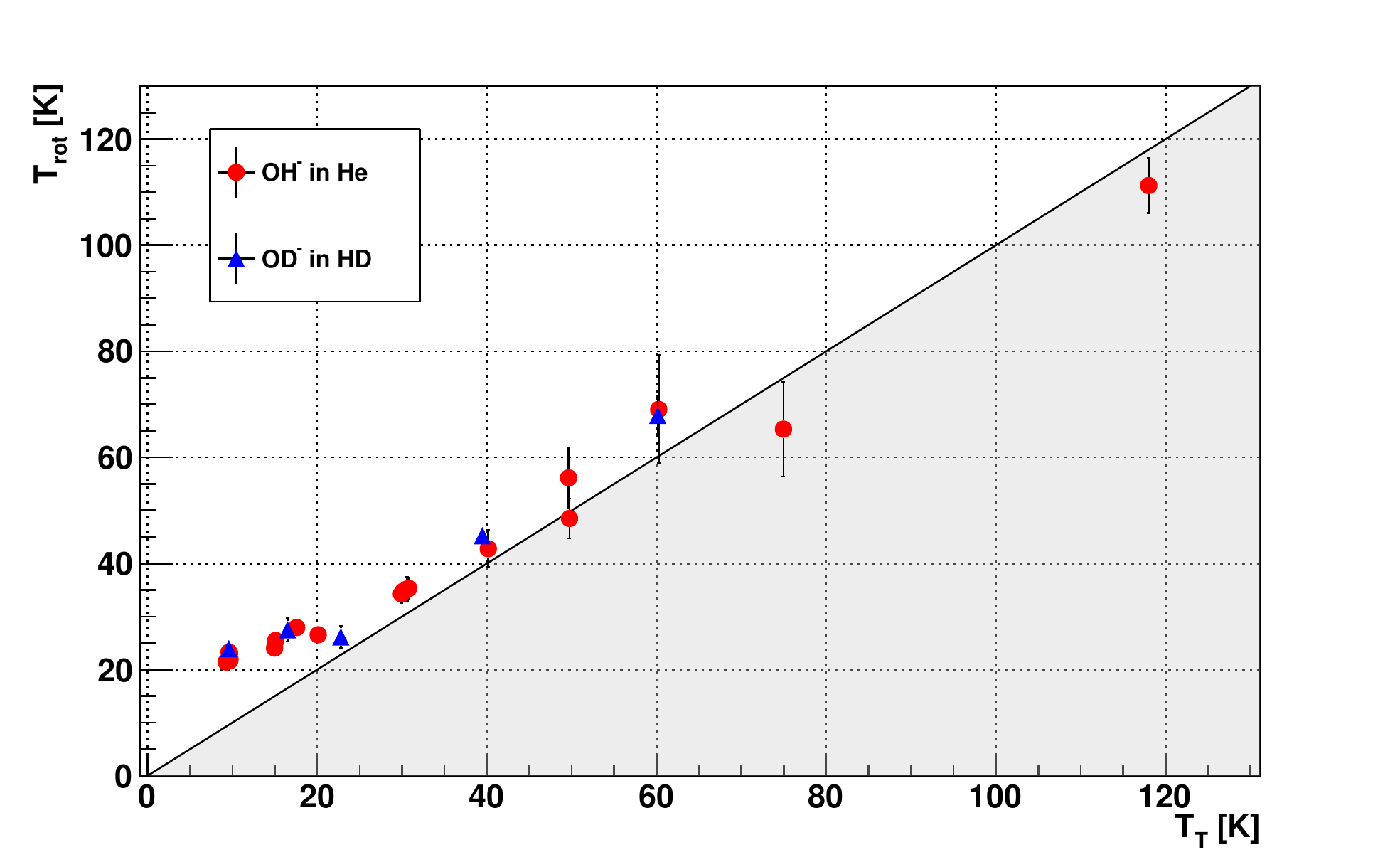}
\caption{Ion rotational temperature $T_{rot}$ vs trap temperature $T_T$ for two different collision partners. The black solid line show the ideal case for $T_{rot} = T_T$}
\label{fig1}
\end{figure}

Experiments were conducted for trap temperatures between $T_T=9$\,K and 120\,K. For each selected temperature $T_T$, internal state thermometry measurements were done to obtain the rotational temperature $T_{rot}$ of OH$^-$ anions in He buffer gas. The resulting rotational temperatures are plotted in Fig. \ref{fig1} against the trap temperature (red circles). Above 40\,K the rotational temperature follows the trap temperature with a reasonably good agreement. But below 40\,K T$_{rot}$ always remains larger than T$_T$ and levels off at about 25\,K. This represents a clear decoupling of the rotational temperature from the trap wall temperature at temperatures below 25\,K. These results qualitatively confirm the results that have been found earlier by Otto et. al. \cite{otto2013:pccp} and extend them to higher temperatures. The lowest rotational temperatures measured in the present work are slightly below the lowest temperature measured there.

To obtain more insight into this temperature decoupling we have varied the ion-to-neutral mass ratio by moving to a lighter buffer gas. We employ HD because in contrast to normal H$_2$ only the rotational ground state is populated at the temperatures of the experiment. We also moved to deuterated hydroxyl anions, because for OD$^-$ the isotopic exchange reaction with HD is suppressed \cite{mulin2015:pccp}. The two systems, OH$^-$/He and OD$^-$/HD, are described by ion-to-neutral mass ratios of 4.25 and 6, respectively. The rotational temperature measurements for OD$^-$ in HD are shown in Fig.\ \ref{fig1} (blue triangles). The data show that for OD$^-$ in HD very similar temperatures are found. Notably also the same decoupling effect is observed, which is therefore independent of the ion-to-neutral mass ratio.

\begin{figure}
\includegraphics[width= \textwidth]{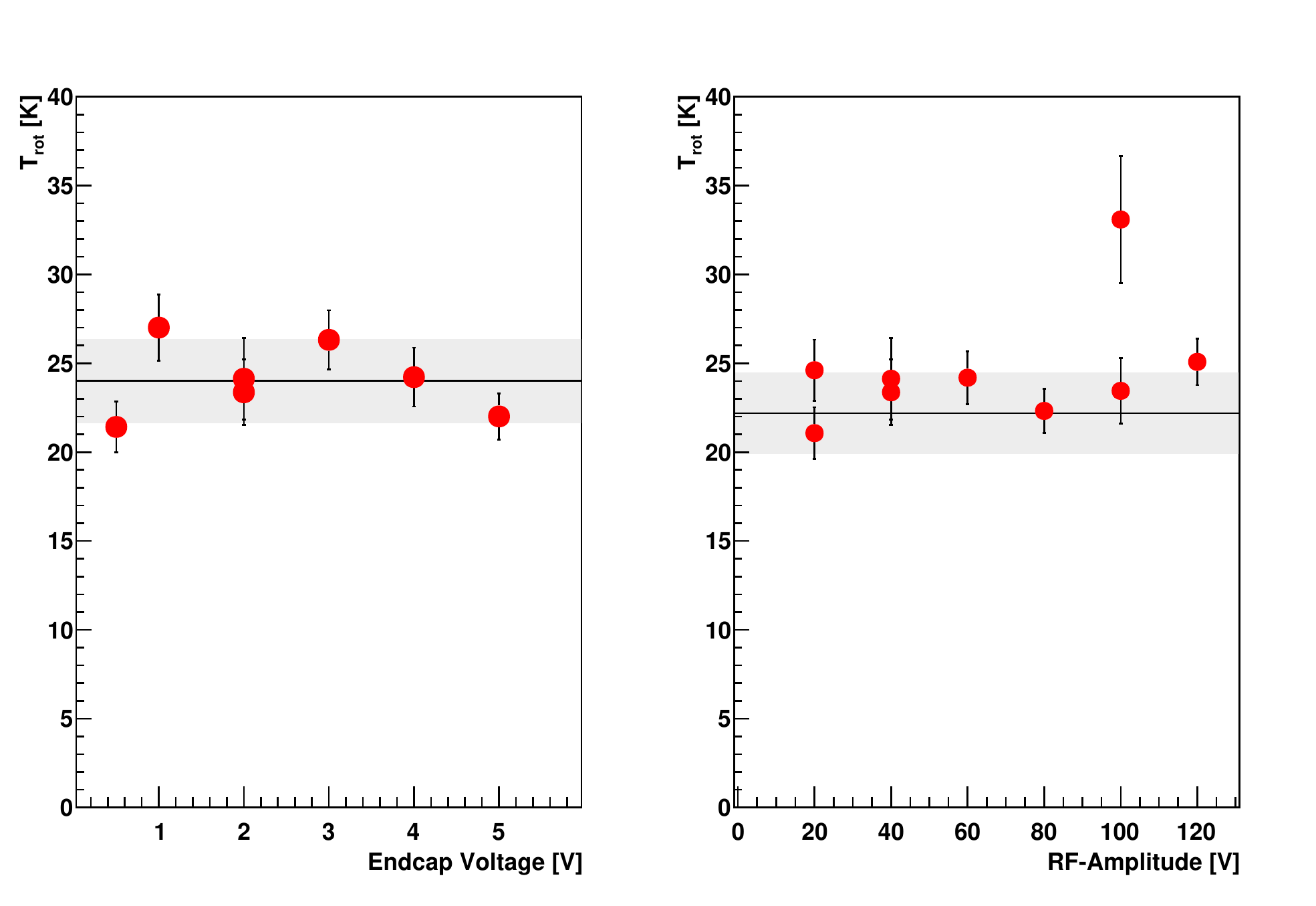}
\caption{Left panel: Measured rotational temperature versus endcap voltage. Right panel: Rotational temperature versus amplitude of the applied alternating field. In both panels the black solid line shows the average rotational temperature.}
\label{fig2}
\end{figure}

In further measurements we have probed the influence of the radial and longitudinal confinement potential of the trap. To this end we varied the endcap voltage and the amplitude of the alternating radiofrequency potential. For each setting the rotational temperature for the OH$^-$/He system was determined, as shown in Fig.\ref {fig2}. In these measurements the trap temperature was fixed at 9\,K. It was found that none of the changes to the trapping potential influenced the rotational temperature.

\section{Discussion}

In the following we discuss different assumptions that could explain the measured increased rotational temperatures at trap temperatures below 25\,K.

\subsection{Radiofrequency heating}

A plausible cause for the increased rotational temperature is collisional heating while the trapped ions are subject to the driven micromotion in the radiofrequency field. This radiofrequency heating is a consequence of the dephasing of the ions' micromotion after elastic collisions with buffer gas atoms. It is strongly dependent on the mass ratio of the colliding partners and the multipole order of the trapping potential \cite{asvany2009:ijm}.

Interestingly, Fig.\ \ref{fig1} clearly shows that the dependence of the rotational temperature on the trap wall temperature is not dependent on the ion-to-buffer gas mass ratio. Fig.\ \ref{fig2} depicts that under variation of the trapping potentials the rotational temperature remains constant. This is contrasted by the fact that increased endcap potentials increase the radially deconfining potential, which pushes the trapped ions closer to the radiofrequency electrodes and should lead to stronger radiofrequency heating. Furthermore, simulations show that radiofrequency heating increases the ions' translational temperature approximately proportional to the buffer gas temperature \cite{asvany2009:ijm,wester2009:jpb}. A decoupling at low temperatures as observed experimentally is therefore not expected. This is supported by further simulations of the velocity distribution of the trapped ions, which we have performed using the specific radiofrequency and endcap potentials employed in our ion trap. All this suggests that standard radiofrequency heating is not responsible for the observed decoupling of temperatures.

\subsection{Buffer gas thermalisation}

Buffer gas cooling of trapped ions is based on the pre-condition that the buffer gas thermalises well with the cryogenic walls of the trap. The wall temperatures are measured with two diodes as described in section \ref{exp:setup}, but the buffer gas temperature is usually difficult to measure. To test if the buffer gas temperature follows the temperature of the trap walls down to 9\,K, we employ a scheme to determine relative temperature changes of the gas using the temperature-dependence of the density according to equation \ref{eq4}.

When adjusting a constant buffer gas flow into the trap this will lead to an equilibrium with a constant pressure measured using the capacitive gauge. As the gauge is held at ambient temperature, its pressure reading is independent of the actual buffer gas temperature in the trap. To probe the temperature-dependent gas density inside the trap we therefore use a different approach. We measure the mean free path of ions moving in a single pass through the ion trap. For ions with a few electronvolt kinetic energy the mean free path scales inversely proportional to the gas density and thus proportional to the square-root of the gas temperature. Decreasing $T_T$ should therefore lead to more deflecting collisions when ions pass through the trap and thus to a lower transmitted ion signal. For small ion losses the mean free path scales linearly with the loss fraction. 

\begin{figure}
\includegraphics[width= \textwidth]{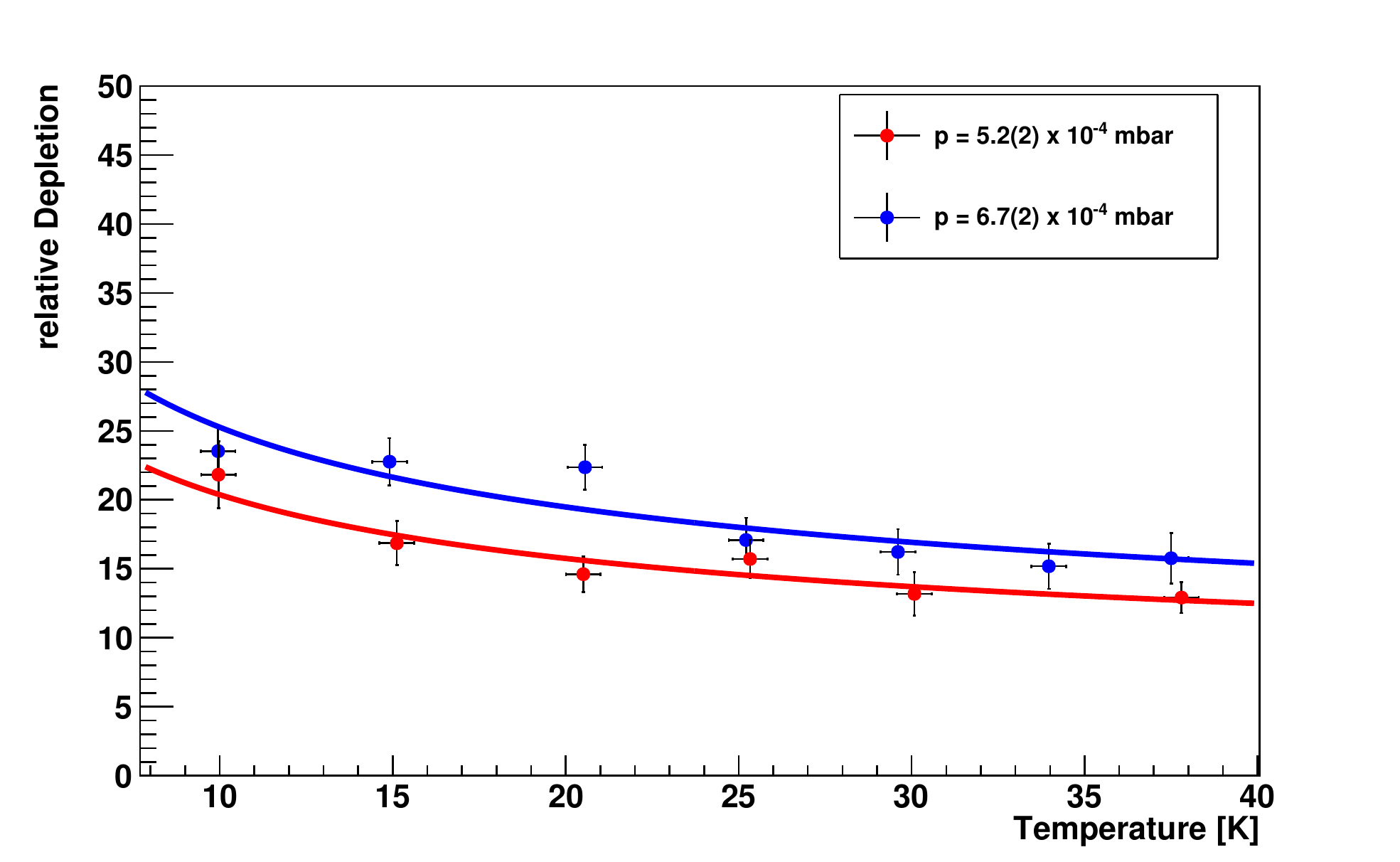}
\caption{Relative depletion of the ion signal at different trap temperatures. The blue points represents measurements at high pressure and the red ones at lower pressures. The solid lines show the fit results of the function according to equation \ref{eq4}.}
\label{fig3}
\end{figure}

We have measured ion transmission signals with and without helium buffer gas using using H$^-$, because the probability for scattering-induced ion losses are largest for light ions. Fig.\ \ref{fig3} shows the the relative change in ion signal as a function of $T_T$. The measurement was repeated for two different settings of the gas flow and thus the capacitive gauge pressure reading. A temperature-dependence is observed in both traces, which shows that the buffer gas cools further down in the temperature regime below 25\,K. The data can be fit reasonably well within the signal-to-noise ratio with the $\frac{1}{\sqrt{T_T}}$-dependence in accordance with equation \ref{eq4}. This suggests that the buffer gas indeed thermalises with the trap walls and can not account for the increased rotational temperature of the trapped ions.

\subsection{Blackbody excitation}

As mentioned before the thermal shield has two openings, where not only warm gas can enter the trap but also radiation. Thus, blackbody radiation from the surrounding room temperature vacuum chamber can enter the trap and influence the internal thermal distribution. The Einstein coefficients $A$ and $B$ for emission and absorption are commonly used to describe the strength of optical transitions, For OH$^-$ the fundamental rotational transition is described by $A\approx3 \cdot 10^{-3}$/s and $B \rho(\omega) \approx 4 \cdot 10^{-2}$/s for a surrounding blackbody radiation field $\rho(\omega)$ at 295\,K \cite{leroy:misc}. For the $J=1$ to 2 transition this increases to roughly $B \rho(\omega) \approx 3 \cdot 10^{-1}$/s. For the fundamental vibrational transition coefficients $A \approx 150$/s and $B \rho(\omega) \approx 10^{-5}$/s are obtained. In the ion trap the blackbody excitation rates are much smaller than the given values, because the room temperature radiation field interacts only via a small solid angle (less than 10\% of the full solid angle) with the trapped ion cloud.

The blackbody coupling rates have to be compared to the inelastic collision rates with helium \cite{hauser2015:natp}. At the densities of the present experiment these collision rates exceed 10$^2$/s. Thus inelastic collision rates are orders of magnitude faster than blackbody coupling rates. Radiative heating is therefore excluded as an influence on the rotational temperature of the trapped ions.

\subsection{Collisions with room temperature gas}

Heating of the trapped ions may also occur by ambient residual gas that enters the cryogenic trap through the endcap electrodes. In the molecular flow regime this room temperature gas travels into the trap without colliding with the cold buffer gas streaming out. Ions colliding with room temperature gas can be excited to higher rotational levels and thereby change the thermal distribution of the internal degrees of freedom for the given buffer gas temperature. To test the importance of this effect, buffer gas was not injected into the trap directly, but was instead inserted into the surrounding vacuum chamber. The helium atoms then enter the trap, thermalise and act as buffer gas for the trapped ions. This approach maximizes the relative collision rate of the trapped ions with room temperature gas compared to gas that has already thermalised inside the trap enclosure. 

\begin{figure}
\includegraphics[width= \textwidth]{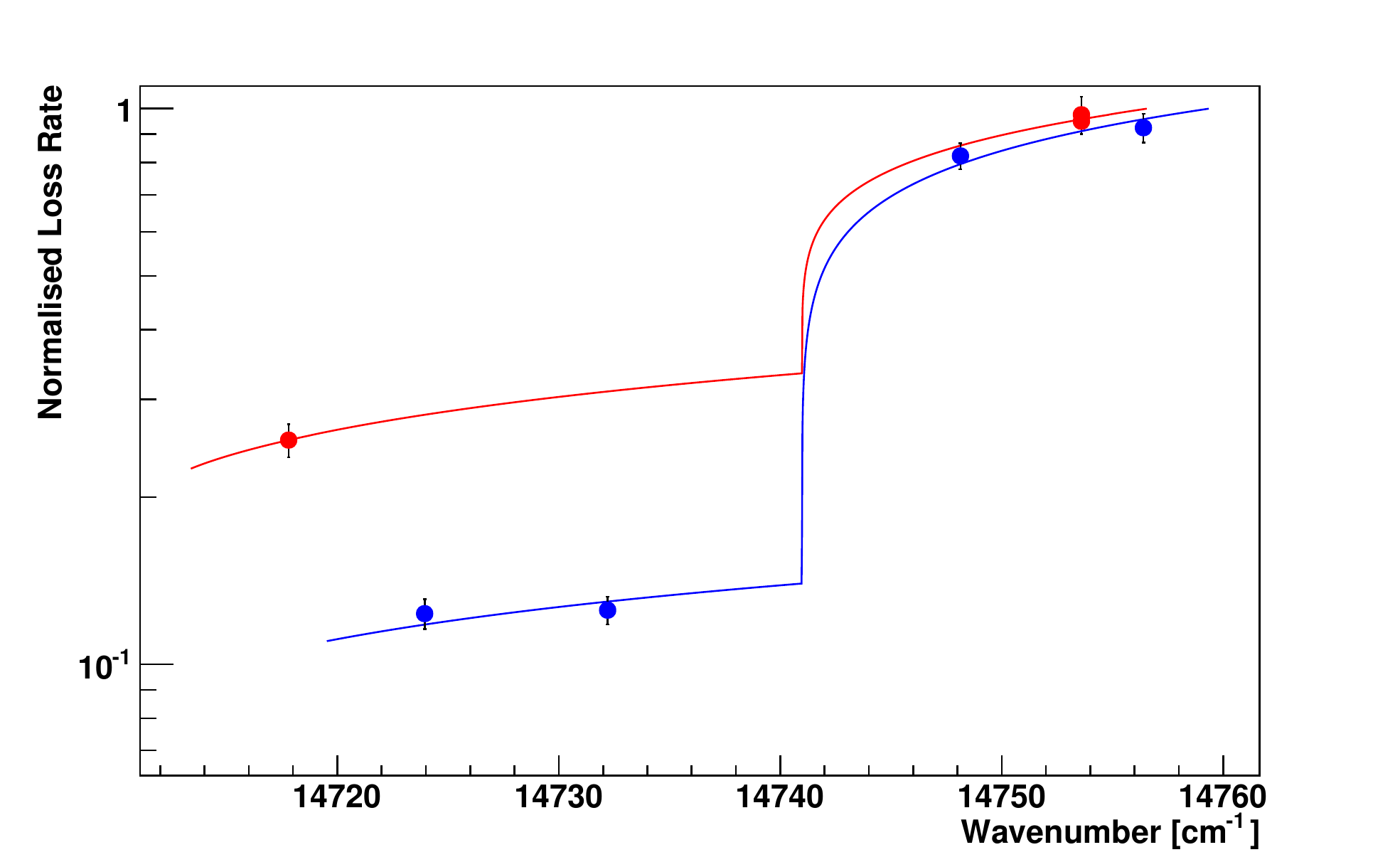}
\caption{Normalized photodetachment loss rate for buffer gas injected into the thermal shield (blue dots) and injected within the surrounding vacuum chamber (red dots). The blue dots show collisions with cold buffer gas and the red ones with warm buffer gas. The lines depict the fit, according to equation \ref{eq1},  to obtain the rotational temperature.}
\label{fig4}
\end{figure}

Fig.\ \ref{fig4} shows photodetachment loss rate measurements using regular buffer gas (blue points) and with buffer gas that only enters the trap through the endcaps (red points). The trap walls were held at 9\,K in both cases. Both data sets are fit to obtain the rotational temperature. The blue data are fit by a rotational temperature of $21.6 \pm 0.6$\,K. For the red data the fit results in a rotational temperature of $36.8 \pm 2.8$\,K. As expected, the influx of room temperature buffer gas increases the rotational temperature. However, the rotational temperature that is reached in steady-state is much below room temperature. This shows that the room temperature gas thermalises efficiently once confined insight the trap due to collisions with the cold trap walls.

As soon as buffer gas is passed directly into the trap the influence of the warm gas can be neglected. This is due to the fact that the relative fraction of collisions with cold buffer gas increases in this case by several orders of magnitude. Rate equation simulations that we have performed for the competition of heating and cooling collisions support this. Thus, also the heating rate introduced by warm buffer gas can not explain the increased rotational temperatures.

\section{Conclusions and Outlook}

Photodetachment experiments show a decoupling of the rotational temperature of the ions from the temperature of the thermal shield of the trap below 40\,K, with $T_{rot}$ leveling off at values of approximately 25\,K. This indicates a breakdown of thermalisation due to potential heating effects that compete with the cooling process. The smallest rotational temperature is independent of the ion buffer gas mass ratio and the confinement potential of the trap, which suggests radiofrequency heating is not the dominant source for rotational heating. Relative ion depletion measurements show a correlation with the temperature of the thermal shield, indicating that the buffer gas does thermalise well with the trap walls. Our analysis shows that blackbody radiation and room temperature gas entering the trap have a small, but quantitatively insufficient heating effect to explain the decoupling of the temperatures. Thus, at present we can not quantitatively explain the source of the rotational heating that leads to the decoupling of the rotational and the trap wall temperature. 

The higher rotational temperature of OH$^-$ compared to the buffer gas temperature may suggest that the translational kinetic temperature of the ions is higher than both other temperatures. In this case the mass-weighted average of the buffer gas temperature and the kinetic temperature, which yields the center-of-mass temperature in the OH$^-$-helium collision system, could correspond to the rotational temperature. To test this experimentally, direct Doppler-resolved spectroscopy is required. Work in this direction is currently in progress in our laboratory.

\section{Acknowledgment}

We thank Stephan Schlemmer for fruitful discussions and for suggesting the buffer gas density measurements. This work has been supported by the European Research Council under ERC Grant Agreement No. 279898. E.S.E. acknowledges support from the Fond National de la Recherche Luxembourg (Grant No. 6019121).

\section*{References}

\providecommand{\noopsort}[1]{}\providecommand{\singleletter}[1]{#1}%

\newpage

\end{document}